
\magnification=\magstep1

\def\etal{{\it et~al.}}

\def\mo{$M_\odot$}
\def\lo{$L_\odot\,$}
\def\mj{$\,M_J\,$}
\def\mp{$\,M_p$}

\def\lbol{$L_{bol}\,$}
\def\mstar{$M_{\ast}$}
\def\Dwa{$\,$\uppercase\expandafter{\romannumeral5}$\,$}

\newcount\sss
\sss=0
\def\super{\advance\sss by 1 $\!^{\number\sss}$}
\def\supe#1{$^{#1}$}

\font\tinyrm=cmr8 at 8truept
\font\tinybold=cmbx8 at 8truept
\font\tinyit=cmti8 at 8truept
\def\tiny{\let\rm=\tinyrm\let\bf=\tinybold\let\it=\tinyit}

\font\smallrm=cmr10 at 10truept
\font\smallbold=cmbx10 at 10truept
\font\smallit=cmti10 at 10truept
\def\small{\let\rm=\smallrm\let\bf=\smallbold\let\it=\smallit}

\font\normalrm=cmr12 at 12truept
\font\normalbold=cmbx12 at 12truept
\font\normalit=cmti12 at 12truept
\def\normal{\let\rm=\normalrm\let\bf=\normalbold\let\it=\normalit}

\font\bigrm=cmr10 scaled 1440
\font\bigbold=cmbx10 scaled 1440
\font\bigit=cmti10 scaled 1440
\def\big{\let\rm=\bigrm\let\bf=\bigbold\let\it=\bigit}

\baselineskip=14truept

\leftline{{\bf Next-Generation Telescopes Can Detect Extra-Solar Giant
Planets}}
\bigskip
\small
\leftline{{\bf A. Burrows$^{\ast}$, D. Saumon$^{\dag}$, T.
Guillot$^{{\ddag}{\dag}}$,
W. B. Hubbard$^{\dag}$, and J. I. Lunine$^{\dag}$}}
\smallskip
\baselineskip=14truept
\leftline{$^{\ast}$Departments of Physics and Astronomy, University of Arizona,
Tucson, AZ 85721 USA}
\leftline{$^{\dag}$Department of Planetary Sciences, University of Arizona,
Tucson, AZ 85721 USA}
\leftline{$^{\ddag}$Observatoire de la C\^ote d'Azur, CNRS/URA 1362, BP 229,
06304 Nice Cedex 4, France}
\baselineskip=24truept
\smallskip
\hrule
\bigskip
\normal

{\bf \noindent Interest among astronomers in the detection of extra-solar
planets is
accelerating with the growing realization that it may soon be technically
feasible.\supe{1}
The ongoing renaissance in telescope construction\supe{2,3}
and the anticipated launches of new space platforms\supe{4,5,6}
are encouraging many scientists to review and improve the means by which
planets can be
discovered.\supe{7,8,9,10,11}
The direct detection of the light from a distant planet would be the most
compelling means of
discovery and to gauge the feasibility of various search strategies,
astronomers have traditionally
used the current Jupiter as a benchmark planet.  However, in principle,
extra-solar
giant planets (EGPs) can have a
wide range of masses and ages,\supe{12}  and, hence, can be significantly
brighter than Jupiter.
Furthermore, the maximum mass a planet can have is
not known from first principles,\supe{13}
and observations will be needed to determine it.  We predict the
optical and infrared fluxes of
EGPs with masses from 0.3 through 15 Jupiter masses and ages from 10$^7$
through $5 \times 10^{9}$ years that
searches in the next few years may reveal.}

EGPs will radiate in the optical by reflection and in the
infrared by the thermal emission of both absorbed stellar light and the
planet's
own internal energy.
To calculate their cooling curves, we used the Henyey code previously
constructed to study
brown dwarfs and M dwarfs.\supe{14,15} Below effective temperatures ($T_{eff}$)
of 600 K, we employed the atmospheres
of Graboske \etal,\supe{16} who included opacities due to water, methane,
ammonia,
and collision-induced absorption by H$_{2}$ and He.  The gravity dependence of
the EGP atmospheres was
handled as in Hubbard.\supe{17}  Above $T_{eff}=600\,$K,  we used the X model
of reference 15.
The two prescriptions were interpolated in the overlap region. We employed the
hydrogen/helium equation of
state of Saumon \& Chabrier\supe{18,19}
and ignored rotation
and the possible presence of an ice/rock core.\supe{20,21}
The EGPs were assumed to be fully convective at all times.
We included the effects of ``insolation'' by a central star of mass \mstar$\,$
and considered
semi-major axes ($a$) between 2.5 A.U. and 20 A.U.  Giant
planets may form preferentially near 5 A.U.,\supe{22} but a range of $a$'s can
not be excluded.
We assumed that the Bond albedo of an EGP is that of Jupiter (0.35).\supe{23}
For this study, we evolved EGPs with masses ($M_p$) from 0.3\mj\thinspace(the
mass of Saturn)
through 15\mj (\mj $\cong1.9\times10^{30}$ g, where \mj$\,$ is the mass of
Jupiter).
Whether a 15\mj object is a planet or a brown dwarf is largely a semantic
issue, though one might
distinguish gas giants and brown dwarfs by their
mode of formation (e.g. in a disk or ``directly'').  Physically, compact
hydrogen-rich objects with masses from 0.00025 \mo$\,$ through
0.25 \mo$\,$ form a continuum.  However, our EGPs
above $\sim$13\mj$\,$ do burn ``primordial'' deuterium for up to 10$^{8}$
years. Note that any search for
giant planets will perforce be even more capable at discovering brown dwarfs.

The evolution of the bolometric luminosity (\lbol) of our suite of EGPs
orbiting 5.2 A.U. from a G2\Dwa star
is depicted in Figure 1.  One is
struck immediately by the high \lbol 's for early ages and high masses.
That a young ``Jupiter'' or ``Saturn'' will be bright has
been known for some time,\supe{16,17,20,21,24,25,26} but ours are the first
detailed calculations for
objects with \mp $\ >$ \mj and ages, $t$, greater than 10$^{7}$ years.
Below about 10\mj, \lbol$\,$ is very roughly proportional
to $M_p^{\alpha}\!/t^{\beta}$, where $1.6 \le \alpha \le 2.1$  and $1.0 \le
\beta \le 1.3$. An EGP with a mass of 2\mj at age
10$^{7}$ years is two thousand times brighter than the current Jupiter (and its
$T_{eff}$ is $\sim$700 K).
At the age of the Pleiades ($\sim7\times10^{7}$ years), such an EGP would be
$\sim 200$ times brighter
(with $T_{eff}$ $\sim 420\,$K) and at the age of the Hyades
($\sim6\times10^{8}$ years) it would be $\sim 18$ times
brighter (with $T_{eff}$ $\sim 235\,$K).  The measured  \lbol and $T_{eff}$ of
Jupiter are
$2.186\pm0.022\times10^{-9}$ \lo and  $124.4\pm0.3\,$K, respectively.\supe{27}
At an age of $4.55\,$Gyr, our
model of Jupiter has a luminosity of $2.35\times10^{-9}$ \lo and an effective
temperature of $122\,$K.

A complete discussion of our results is deferred to a later paper (Saumon \etal
, in preparation).
However, a few ``facts'' will serve to illustrate the character of massive,
young EGPs.
Since the fluxes shortward
of $10\, \mu $m (the N band) are generally on or near the Wien tail of the EGP
spectrum, the
fluxes in the near- and mid-infrared spectral bands increase even faster
with mass and youth than \lbol .  In particular, assuming that the emission is
Planckian, that the orbital separation
is 5.2 A.U., and
that \mstar$\,$ equals 1.0$\,M_\odot$, Jupiter's N band flux would be $\sim
8000$ times higher at age
$10^{7}$ years than it is now.  At the age of our solar system,  a 2\mj EGP and
a 5\mj EGP would be $\sim6$  and $\sim90$
times  brighter in the N band than the current Jupiter.
Furthermore,
in the M band ($\sim5\,\mu $m), a 2\mj\ EGP would be $\sim\!60,\!000$ times
brighter at $10^{7}$
years than the current Jupiter, but at $10^{9}$ years ``only'' $\sim2.5$ times
brighter than a coeval Jupiter.  At the
age of the Hyades, Saturn would be as bright as the current Jupiter.  The
fluxes at Earth
due to the thermal emissions shortward of $10\,\mu $m of EGPs in the Pleiades
($D\sim125\,$ parsecs)
would be greater than those from EGPs in the Hyades ($D\sim45$ parsecs),
despite the latter's
relative proximity, because the Pleiads are younger (and, hence, at higher
$T_{eff}$).

In this paper, we focus on the crucial question of the brightness of EGPs as a
function of age and mass
to aid in the development of search strategies.
However,  the detection of EGPs requires telescopes with both high sensitivity
{\it and}
high angular resolution.  The latter is necessary to discriminate the planet
from the star and can be
compromised by the presence of
scattered light in the
telescope optics.
Nevertheless, it is expected that the Large Binocular Telescope\supe{2} (LBT),
the Near Infrared Camera and
Multi-Object Spectrometer\supe{4} (NICMOS), and the Space InfraRed Telescope
Facility\supe{5} (SIRTF) will have both the sensitivity and
the angular resolution (see caption for Figure 3)
to discover EGPs with a variety of realistic combinations of $a$ and $D$.
It is not our purpose here to discuss various detection strategies, nor
to explore the consequences of every combination of \mp, \mstar, $a$, $D$, and
telescope.
Rather, in Figures 2 and 3, we compare
theoretically predicted fluxes for a few representative values of $a$ and $D$
and various EGP ages and masses
with the flux sensitivities of various ground- and space-based telescopes
currently being developed.

Figure 2 depicts the flux in Janskys versus wavelength at 10 parsecs
for a 1\mj and a 5\mj$\,$ EGP that are 5.2 A.U. from a G2\Dwa
star (a solar analog), at times between $10^7$ and $5\times10^{9}$ years.  We
have made
the assumption that the thermal emissions are blackbody and have included the
reflected light.
Also shown on Figure 2 are the $5\sigma$ point-source sensitivities at various
wavelengths for
SIRTF,\supe{4} the LBT,\supe{2} the upgraded ``Multiple Mirror''
Telescope\supe{2} (MMT), Gemini,\supe{3}
the Stratospheric Observatory For Infrared Astronomy\supe{28} (SOFIA), and the
three NICMOS cameras.\supe{4}
As is indicated in Figure 2, the LBT and NICMOS have the flux sensitivity to
see at 10 parsecs the reflected light of
such EGPs at any age.  At the diffraction limit, these instruments
will also have the requisite angular resolution.
At 10 parsecs, SIRTF has the flux sensitivity between $5\,\mu $m and $10\,\mu
$m
to detect the thermal emissions
of both a 5\mj EGP, for ages less than $10^9$ years,
and a Jupiter at 10 A.U., for ages less
than $10^{8}$ years.
Figure 3 shows the fluxes from EGPs with masses between 0.3\mj and 5\mj
orbiting at 10 A.U.
around an A0\Dwa star ({\it e.g.,} a Vega analog) at a distance of 10 parsecs
from the Earth.  Its age is $2\times10^8$
years, just shy of its main-sequence lifetime.  It can be seen that the
reflected light from
all EGPs at 10 A.U. around such a bright star will be detectable by the LBT,
the MMT, and NICMOS, since all
these platforms will achieve angular resolutions well below 1$^{\prime\prime}$.
Furthermore,
the SIRTF sensitivities in the mid-IR would be just adequate to see even a
Saturn-mass object
around an A0\Dwa star at this age, distance,
and separation.
Though SIRTF may not achieve the necessary 1$^{\prime\prime}$ performance,
it should have the angular resolution to discover similar EGPs
at somewhat smaller $D$'s and larger $a$'s.

Interestingly, NICMOS is sensitive enough to detect any widely-separated EGP
with a mass greater than 6\mj
around any
main-sequence star as far away as the Pleiades, while SIRTF has the sensitivity
to detect any ``free-floating  EGPs'' in the Pleiades with a mass greater than
4\mj.  This is because the internal energy
of a young EGP alone is adequate to power it in the relevant spectral bands.
None of the anticipated telescope systems will have the angular resolution to
probe for an EGP in the Pleiades, unless it is many tens of A.U. from its
central star.

Clearly, searches around early main sequence stars or in very young stellar
systems will select for
the youngest and most massive planets.  Crude models of giant planet formation
around a solar-mass
star permit objects up to 10 Jupiter masses to form.\supe{12}  The most massive
planet in our solar system is Jupiter
and why there are none more massive is not understood. The lifetime  of our
gaseous proto-planetary disk
may have been short\supe{29} or the disk may have been tidally truncated by the
growing planet itself.\supe{13}
Good statistics on the population of EGPs will better constrain
proto-planetary disk processes.
The absence of a Jupiter-class planet in a planetary system would
imply a very different population of cometary-sized planetesimals than exists
in our own solar system,
and this may have important implications for the origin, evolution, and
survival of life on
rocky ``terrestrial'' planets.\supe{30} Only sensitive and systematic searches
such as those we anticipate over the
next decade will directly address these important issues of planet formation.

\medskip
\hrule
\bigskip
\vfill\eject
\centerline{\bf References}
\bigskip
\item{1.} TOPS: Toward Other Planetary Systems (NASA Solar System Exploration
Division, Washington, D.C. 1992).
\item{2.} Angel, R.\  {\it Nature} {\bf 368}, 203-207 (1994) (LBT and MMT).
\item{3.} Mountain, M., R. Kurz, R., \& Oschmann, J.\  in {\it The Gemini 8-m
Telescope Projects,
S.P.I.E. Proceedings on Advanced Technology Optical Telescopes V} {\bf 2199},
p. 41 (1994).
\item{4.} Thompson, R.\ {\it Space Science Reviews} {\bf 61}, 69-63 (1992)
(NICMOS).
\item{5.} Erickson, E. F. \& Werner, M. W.\ {\it Space Science Reviews} {\bf
61}, 95-98 (1992) (SIRTF).
\item{6.} Benvenuti, P. \etal\ in {\it ESA's Report to the 30th COSPAR Meeting,
ESA SP-1169, Paris},  p. 75 (1994) (ISO).
\item{7.} Gatewood, G. D. \ {\it Astron.\ J.} {\bf 94}, 213 (1987).
\item{8.} Reasenberg, R. D.~\etal\ {\it Astron.\ J.} {\bf 96}, 1731 (1988).
\item{9.} Walker, G.A.H.~\etal\ {\it Icarus}, in press (1995).
\item{10.} Mao, S. \& Paczynski, B.\ {\it Astrophys.\ J.} {\bf 374}, L37
(1991).
\item{11.} Borucki, W. J. \& Summers, A. L.\ {\it Icarus} {\bf 58}, 121 (1984).
\item{12.} Podolak, M., Hubbard, W. B., \& Pollack, J. B.\ in {\it Protostars
and Planets III},
eds. E. Levy \& J. I. Lunine, University of Arizona Press, P. 1109 (1993).
\item{13.} Lin, D.N.C. \& Papaloizou, J.\ {\it M.N.R.A.S.} {\bf 186}, 799
(1979).
\item{14.} Burrows, A., Hubbard, W. B., \& Lunine, J. I.\ {\it Astrophys.\ J.}
{\bf345}, 939 (1989).
\item{15.} Burrows, A., Hubbard, W. B., Saumon, D., \& Lunine, J. I.\ {\it
Astrophys.\ J.} {\bf 406}, 158 (1993).
\item{16.} Graboske, H. C., Pollack, J. B., Grossman, A. S., \& Olness, R. J.\
{\it Astrophys.\ J.} {\bf 199}, 265-281 (1975).
\item{17.} Hubbard, W. B.\ {\it Icarus} {\bf 30}, 305 (1977).
\item{18.} Saumon, D. \& Chabrier, G.\ {\it Phys. Rev. A} {\bf 44}, 5122-5141
(1991).
\item{19.} Saumon, D. \& Chabrier, G.\ {\it Phys. Rev. A} {\bf 46}, 2084-2100
(1992).
\item{20.} Pollack, J. B.\ {\it Ann. Rev. Astron. Astrophys.} {\bf 22}, 389
(1984).
\item{21.} Bodenheimer, P. \& Pollack, J. B.\ {\it Icarus} {\bf 67}, 391
(1986).
\item{22.} Boss, A. P.\ {\it Science} {\bf 267}, 360 (1995).
\item{23.} Conrath, R. A., Hanel, R. A., \& Samuelson, R. E.\ in {\it Origin
and Evolution of Planetary
and Satellite Atmospheres}, University of Arizona Press, p. 513 (1989).
\item{24.} Saumon, D., Hubbard, W. H., Chabrier, G., \& Lunine, J. I.\ {\it
Astrophys.\ J.} {\bf 391}, 827 (1992)
\item{25.} Guillot, T., Chabrier, G., Gautier, D., \& Morel, P.\ {\it
Astrophys.\ J.}, in press (1995).
\item{26.} Black, D. C.\ {\it Icarus} {\bf 43}, 293-301 (1980)
\item{27.} Pearl, J. C. \& Conrath, R. A.\ {\it J. Geophys. Res. Suppl.} {\bf
96}, 18921-18930 (1991).
\item{28.} Erickson, E. F.\ {\it Space Science Reviews} {\bf 61}, 61-68 (1992)
(SOFIA).
\item{29.} Zuckermann, B., Forveille, T., \& Kastner, J. H.\ {\it Nature} {\bf
373}, 494-496 (1995).
\item{30.} Wetherill, G. W.\ {\it Lunar Planet. Sci. Conf. XXIV}, 1511 (1993).
\item{31.} Dreiling, L. A. \& Bell, R. A.\ {\it Astrophys.\ J.} {\bf 241},
736-758 (1980)
\bigskip
{\noindent Acknowledgment: The authors would like to thank Roger Angel, Nick
Woolf, George Rieke, Frank Low,
Peter Eisenhardt, Dave Sandler,  and Glenn Schneider for many fruitful
discussions on detector technology
and for providing us with instrument specifications and  NASA,
the US NSF, the Hubble Fellowship Program, and the European
Space Agency for financial support.}

\vfill\eject
\baselineskip=22truept
\centerline{FIGURE CAPTIONS}
\bigskip

\item{Figure 1:} Bolometric luminosity (\lbol) in solar units of a suite of
EGPs placed at a
          distance of 5.2 A.U. from a G2\Dwa star
          versus  time ($t$) in Gyr.  The reflected luminosity is not included,
but
          the absorbed component is. At $t \sim 0.2\,$Gyr, the luminosity
          of the 14\mj EGP exceeds that of the 15\mj EGP because of
          late deuterium ignition.
          The data point at $4.55\,$Gyr shows the
          observed luminosity of Jupiter.$^{27}$  The 0.3\mj EGP
          exhibits a strong effect of warming by the G2\Dwa primary star at
          late stages in its evolution.
          Although this model resembles
          Saturn in mass, here it is placed at the distance of Jupiter from
          its primary.
          (The flattening in $L$ vs. $t$ for low masses and great ages is a
          consequence of stellar insolation.)
          The insert shows, on an expanded scale, the comparison
          of our lowest-mass evolutionary trajectories with the present
          Jupiter luminosity.

\item{Figure 2:} Spectral dependence of the flux received at the Earth  from
          extra-solar giant planets (EGPs) orbiting at 5.2 A.U. from a G2\Dwa
          star at 10 parsecs from the Earth (The orbit subtends an angle
          of $\sim$0.5$^{\prime\prime}$).  Objects of 1\mj (top panel)
          and 5\mj (bottom panel) are displayed at the
          following ages: log $t ({\rm yr})=$ 7.0, 7.5, 8.0, 8.5, 9.0,
          9.5, 9.7 (from left to right).  The reflected component of the
          light is essentially independent of the mass and the age of the EGP.
          The EGP is assumed to emit like a blackbody and to reflect incident
          light as a grey body.  Standard photometric bandpasses are shown at
          the top.  Also shown are the design sensitivities of several
          astronomical systems for the detection of point sources with
          a signal-to-noise ratio of 5 in a 1-hour integration (40 minutes
          for NICMOS). These systems are the LBT and
          MMT (solid circles and square, respectively), the three cameras
          of NICMOS (open triangles, 3-pointed stars and solid triangles),
          SIRTF (solid bars), and Gemini and SOFIA (dashed bars).
          The spectrum of a G2\Dwa star was provided by A. Eibl (private
          communication, 1995).
          It should be stressed that while SIRTF is unlikely to have the {\it
          angular resolution} to detect the EGPs of this example, the same EGPs
          at slightly larger separations around a star that is slightly closer
          will be well within its detection envelope (see caption for Figure
3).

\item{Figure 3:} Same as Figure 2, but for EGPs orbiting 10 A.U. from an
          A0\Dwa star whose age is $2.0\times10^{8}$ years and whose distance
is 10 parsecs.
          (The angular separation is $1^{\prime\prime}$.)
          From left to right,
          the curves correspond to masses of 5, 4, 3, 2, 1, 0.5 and 0.3\mj .
          The spectrum of an A0\Dwa star is from Dreiling \& Bell.\supe{31}
          Note that the heating by this star at 10 A.U. is quite significant
          for all lower-mass EGPs and that
          the reflected component of an EGP is $\sim$80 times brighter when it
orbits an A0\Dwa star
          than when it orbits a G2\Dwa star.
          The MMT, the LBT, and NICMOS (on the Hubble Space Telescope),
          with apertures of 6.5 meters, 8.4 meters ($\times 2$), and 2.4
meters, respectively,
          will achieve angular resolutions
          near their respective diffraction limits (R. Angel,\supe{2} N. Woolf,
and G. Schneider, private communications).
          Being a mid-infrared platform
          with a 0.85-meter aperture, SIRTF's angular resolution will, quite
naturally, be poorer than that of the optical and
          near-infrared platforms.
          However, with super-resolution techniques,
          SIRTF should be able to resolve EGPs at
          somewhat larger $a$'s and smaller distances than used for this figure
(F. Low, private communication).

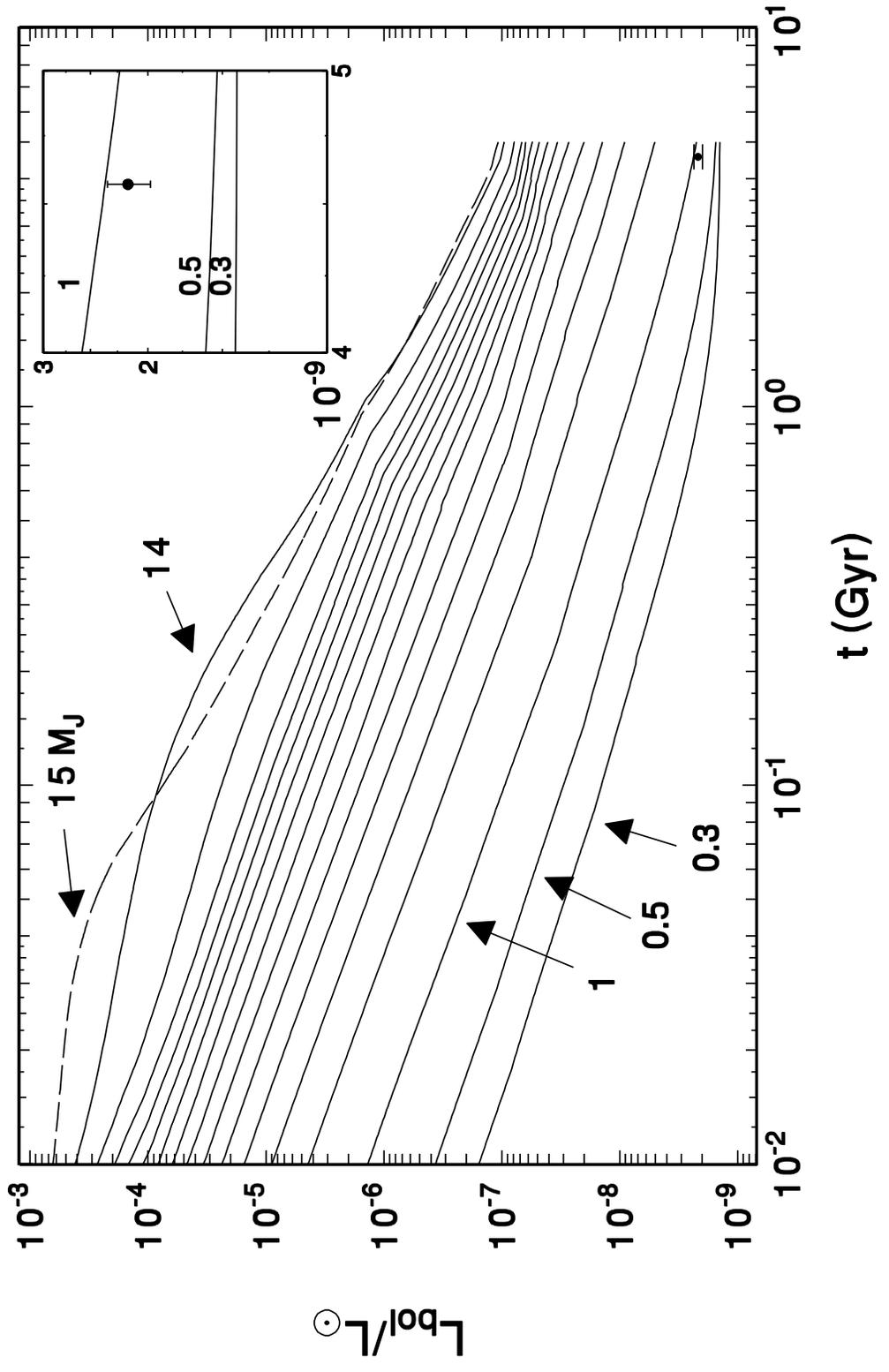
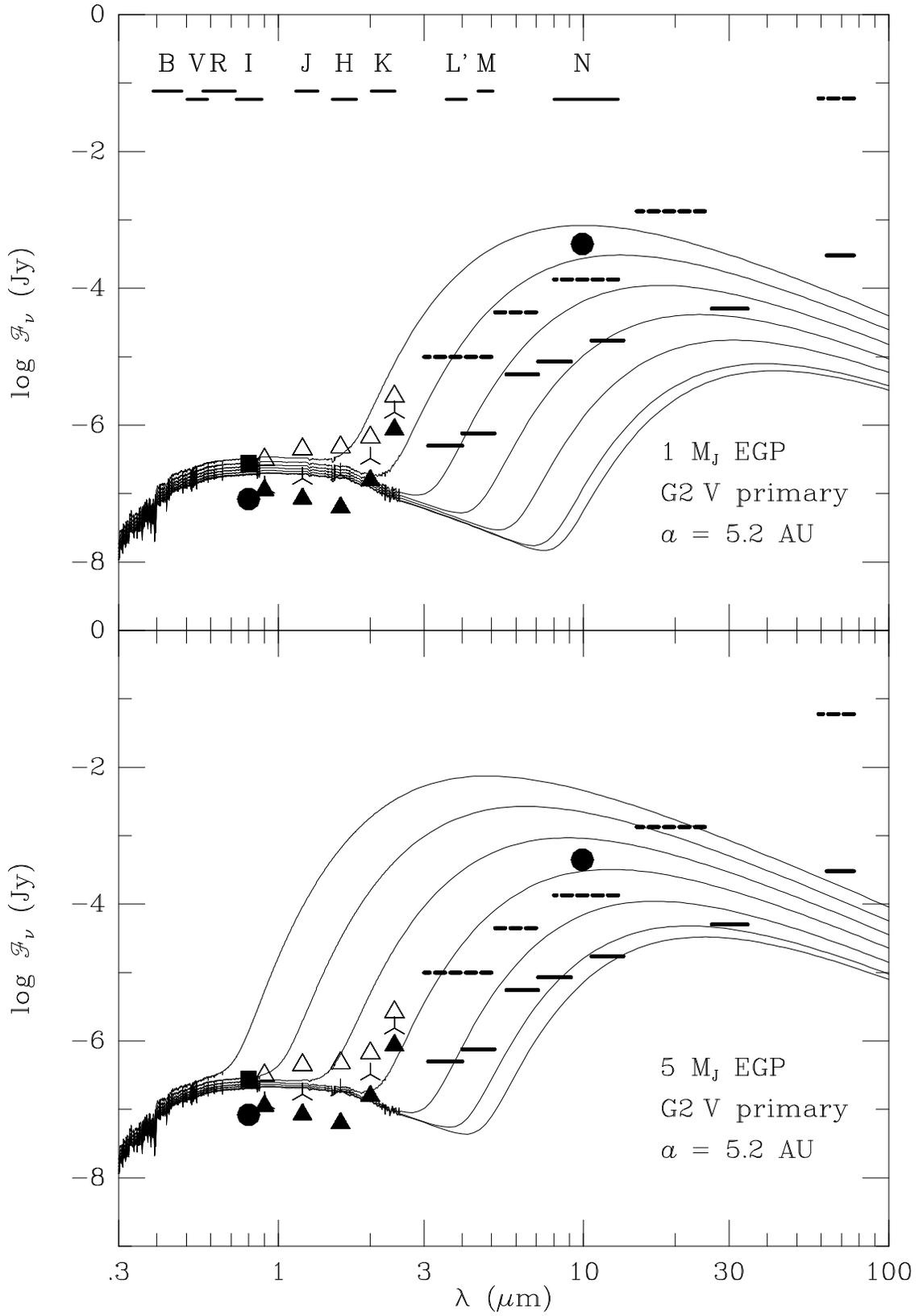
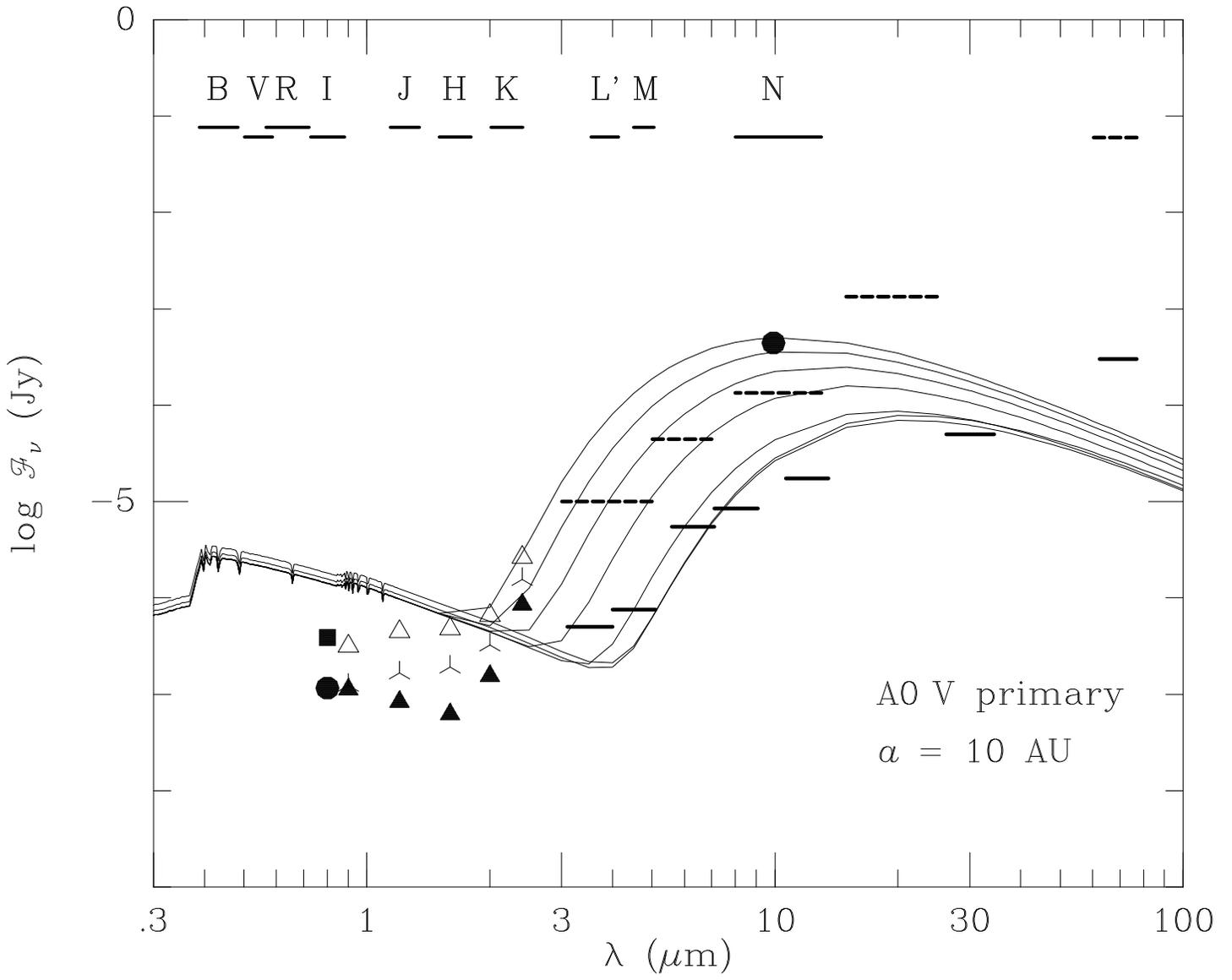
\bye